\documentstyle[preprint,aps,epsfig]{revtex}

\newcommand{\be}{\begin{eqnarray}}
\newcommand{\ee}{\end{eqnarray}}

\newcommand{\e}{\epsilon}

\begin{document}


\preprint{
KAIST-TH-97/19
}

\title{ 
Phenomenological Implications of the Topflavour Model
}
\vspace{0.65in}

\author{
Jong Chul ~Lee\thanks{jclee@chep5.kaist.ac.kr},
Kang Young ~Lee\thanks{kylee@chep5.kaist.ac.kr 
}
and Jae Kwan ~Kim
}
\vspace{.5in}

\address{
Department of Physics, Korea Advanced Institute of Science and Technology\\
Taejon 305 -- 701, Korea}

\date{\today}
\maketitle

\begin{abstract}
\\
We explore phenomenologies of the topflavour model 
for the LEP experiment at $m_{_Z}$ scale and 
the atomic parity violation (APV) experiment in the $Cs$ atoms
at low energies.
Implications of the model on the $Z$ peak data are studied
in terms of the precision variables $\epsilon_i$'s.
We find that the LEP data give more stringent constraints
on the model parameters than the APV data.

\end{abstract}

\pacs{ }

\narrowtext

\section{Introduction}

After the observation of top quark
at the Fermilab $p \bar p$ collider, Tevatron,
influence of the large value of top quark mass 
on the $Z \to b \bar b$ vertex has been drawing much attention.
For instance, the LEP measured quantity
$R_b \equiv \Gamma(Z \to b \bar b)/\Gamma(Z \to \mbox{hadrons})$
has been known to show some inconsistency with the Standard Model (SM)
prediction with such a heavy top.
Following the recent LEP electroweak working group report \cite{LEPEWWG},
the average of LEP (LEP+SLC) data is $R_b = 0.2174 \pm 0.0009$
( $R_b = 0.2170 \pm 0.0009$ ),
which is about 1.8 (1.5) standard deviations higher than the SM prediction.
In fact the newly analyzed value from new data come closer 
to the SM prediction and the experimental situation is being improved.
As a result, the discrepancy of $R_b$ is expected to be 
eliminated in the near future.
In spite of the dilution of $R_b$ problem, however, 
it is still interesting to study
the third generation related phenomenologies
where the new physics manifests.


Extensions of the standard electroweak interactions
with additional SU(2) group has been extensively studied 
in many literatures \cite{susu1,susu2,susu3,malkawi,nandi}.
Depending on the motivations of extensions, 
SU(2)$\times$SU(2) groups are differently coupled to fermions, 
and various models are suggested accordingly.
In this paper we consider the model with additional SU(2) 
acting only on the third generation, 
which has been suggested by several authors \cite{malkawi,nandi}
as a possible solution of the $R_b$ problem.
The third generation undergoes different flavour dynamics 
from the first and second generations
and we expect that this type of model would help us to explain
the discrepancy of $R_b$ which still exists.
As an analogy to the topcolor model \cite{topcolor},
this model is called topflavour model \cite{nandi}.

In order to parametrize new physics effects on the observables
from the LEP experiments, we use the precision
variables $\epsilon_i$'s introduced by Altarelli et al. 
\cite{altarelli1,altarelli2} in this paper.
Because there is no direct evidence for new physics beyond the SM
at LEP until now, the new physics contribution, if it exists,
are thought to be at most comparable with the radiative correction
effects of the SM.
Hence it will be interesting to study the new physics effects
in terms of the precision variables.
We calculate $\epsilon_{1,2,3,b}$ 
with the new LEP data reported by the LEP Electroweak Working Group
\cite{LEPEWWG} in the framework of the topflavour model.
Among four epsilon variables, $\epsilon_b$ has been of particular interest
because it encodes the corrections to the $Z \to b \bar b$ vertex 
and is relevant for our aim.

This model shows a characteristic feature, 
existence of the flavour changing neutral currents (FCNC)
even for the lepton sector. 
We use the experimental bound of FCNC
in the lepton sector combined with the
precision test to result in additional constraints
on the allowed parameter space.

Another important constraint on new physics comes from 
measurements of the atomic parity violation (APV) in heavy atoms.
Deviations from the SM predictions in APV
may be due to the presence of extra neutral gauge bosons
and useful to restrict the mass of such gauge bosons \cite{marciano,apv}.
Here we analyze such deviations in the topflavour model
and compare the results with those of the $Cs$ atom which is accurately
measured recently.

This work is organized as follows:
The model is briefly reviewed in section II.
We calculate $\epsilon_{1,2,3,b}$ in the topflavour model and
with the new LEP data in section III.
The model parameters are constrained 
by the results of the experimental data.
We discuss the FCNC and lepton mixings of the topflavour model
in section IV.
The APV in the topflavour model is analyzed in section V.
Discussions and conclusions are given in section VI.

\section{The Model}

We study the topflavour model 
with the extended electroweak gauge group  
$SU(2)_l \times SU(2)_h \times U(1)_Y $  where 
the first and second generations couple to $SU(2)_l$ 
and the third generation couples to $SU(2)_h$.
The left--handed quarks and leptons in the first and second generations
transform as (2,1,1/3), (2,1,-1) under 
$SU(2)_l \times SU(2)_h \times U(1)_Y $,
and those in the third generation as (1,2,1/3), (1,2,-1)
while right--handed quarks and leptons transform as (1,1,2$Q$)
where $ Q = T_{3l} + T_{3h} + Y/2 $ 
is the electric charge of a fermion.
 
The covariant derivative is given by 
\begin{equation}
 D^{\mu} = \partial^{\mu} + i g_l T^{a}_{l} W^{\mu}_{la} 
          + i g_h T^{a}_{h} W^{\mu}_{ha} 
          + i g^{\prime} \frac{Y}{2} B^{\mu} , 
\end{equation}
where $ T^{a}_{l}$ and $T^{a}_{h}$ denote the $SU(2)_{(l,h)}$ 
generators and $Y$ is the $U(1)$ hypercharge generator. 
Corresponding gauge bosons are $ W^{\mu}_{la} , W^{\mu}_{ha} $
and $B^{\mu}$ with the coupling constants  
$g_{l}, g_{h}$ and $g^{\prime}$ respectively.
The gauge couplings may be written as
\begin{equation}
g_l = \frac{e}{\sin \theta \cos \phi} ,\mbox{~~~~}
g_h = \frac{e}{\sin \theta \sin \phi},\mbox{~~~~} 
g^{\prime} = \frac{e}{\cos \theta} 
\end{equation}
in terms of the weak mixing angle $\theta $ and the new mixing 
angle $\phi$ between $SU(2)_l$ and $SU(2)_h$ defined in eq. (3) below.

The symmetry breaking is accomplished  by the vacuum expectation values 
of two scalar fields  $\Sigma$ and $\Phi$
\begin{eqnarray} 
\langle \Phi \rangle = \left( \begin{array}{c}
                                 0 \\
                                 v/\sqrt{2} 
                               \end{array} 
                         \right) , ~~~~~
\langle \Sigma \rangle =  \left( \begin{array}{cc}
                                   u & 0  \\
                                   0 & u  \\ 
                                 \end{array}
                          \right)~~.
\nonumber
\end{eqnarray}
The scalar field $\Sigma = \sigma + i \vec{\tau} \cdot \vec{\pi} $ 
transforms as (2,2,0) under $ SU(2)_l \times SU(2)_h \times U(1) $
and $\Phi$ as (2,1,1), the latter corresponds to the SM Higgs field.   
Though we do not explicitly describe the potential of Higgs fields here,
it should be suitably chosen to give the correct vacuum.
In the first stage, 
the scalar field $\Sigma$ gets the vacuum expectation value and
breaks $SU(2)_l \times SU(2)_h \times U(1)_Y $ down 
to $ SU(2)_{l+h} \times U(1)_Y $ at the scale $\sim u$.
The remaining symmetry is broken down to $U(1)_{em}$  
by the the vacuum expectation value of $\Phi$ 
at the electroweak scale. 
Since the third generation fermions do not couple to Higgs fields
with this particle contents,
they should get masses via higher dimensional operators.
The different mechanism of mass generation
could be the origin of the heavy masses of the third generation.
Or it would be possible to introduce another Higgs doublet
coupled to the third generations as in the ref. \cite{nandi}.
We do not study the mass generation problem of this model
in details here.


We assume that both $SU(2)$ interactions are perturbative 
and the first symmetry breaking scale is much higher than
the electroweak scale, $ v^2/u^2 \equiv \lambda \ll 1 $.
The value of the mixing angle $\sin \phi$ 
is constrained in order for this model to be perturbative,
$g_{(l,h)}^2/4 \pi < 1$,
such that $ 0.03 < \sin^2 \phi < 0.96 $.

Let us consider the mass eigenstates of gauge bosons. 
It is convenient to write the gauge bosons in the following basis, 
\begin{eqnarray} 
W^{\pm}_{1 \mu} &=& \cos \phi W^{\pm}_{l \mu} 
                 + \sin \phi W^{\pm}_{h \mu}~~,
\nonumber\\ 
W^{\pm}_{2 \mu} &=& -\sin \phi W^{\pm}_{l \mu} + \cos \phi W^{\pm}_{h \mu}~~,
\nonumber\\ 
Z_{1\mu} &=& \cos \theta ( \cos \phi W^{3}_{l \mu} + \sin \phi W^{3}_{h \mu}) 
             - \sin \theta B_{\mu} ~~,
\nonumber\\ 
Z_{2\mu} &=&  -\sin \phi W^{3}_{l \mu} + \cos \phi W^{3}_{h \mu}~~,
\nonumber\\ 
A_{\mu}  &=& \sin \theta ( \cos \phi W^{3}_{l \mu} + \sin \phi W^{3}_{h \mu}) 
            + \cos\theta B_{\mu},
\end{eqnarray}  
where $ W^{\pm}_{(l,h),\mu} = ( W^{1}_{(l,h),\mu} 
                                \mp i W^{2}_{(l,h),\mu}) /\sqrt{2} $.
Up to the order of $\lambda$ the physical states of gauge bosons are 
given by
\begin{eqnarray} 
   \left( \begin{array}{c} 
          W^{\pm}_{\mu} \\
          W^{\prime \pm}_{\mu} 
          \end{array}
   \right)
  &=& 
   \left( \begin{array}{cc} 
      1                            &  \lambda \sin^3 \phi \cos \phi  \\
   - \lambda \sin^3 \phi \cos \phi  & 1  
          \end{array}
   \right)
   \left( \begin{array}{c} 
          W^{\pm}_{1\mu} \\
          W^{\pm}_{2\mu} 
          \end{array}
   \right)~~,
\nonumber \\
   \left( \begin{array}{c} 
          Z_{\mu} \\
          Z^{\prime }_{\mu} 
          \end{array}
   \right)
  &=& 
   \left( \begin{array}{cc} 
      1                      &  \lambda \frac{\sin^3 \phi \cos \phi}
                                             {\cos\theta}  \\
   - \lambda \frac{\sin^3 \phi \cos \phi}{\cos \theta}  & 1  
          \end{array}
   \right)
   \left( \begin{array}{c} 
          Z^0_{1\mu} \\
          Z^0_{2\mu} 
          \end{array}
   \right)~~,
\end{eqnarray} 
with the masses 
\begin{eqnarray}
m^{2}_{W^{\pm}} = m^2_{0} ( 1 - \lambda \sin^4 \phi ), ~~~~ 
m^{2}_{Z} = \frac{m^2_{0}}{\cos^2 \theta}( 1 - \lambda \sin^4 \phi ), 
\nonumber \\
m^{2}_{W^{\prime \pm}} = m^{2}_{Z^{\prime}} 
= m^2_{0} \left(
 \frac{1}{\lambda \sin^2 \phi \cos^2 \phi} 
+ \frac{\sin^2 \phi}{\cos^2 \phi} 
\right) 
\end{eqnarray} 
where $m_{0} = e v /(2 \sin \theta )$ is the tree level mass of the 
$W$ boson in the SM.
In terms of mass eigenstates of gauge bosons 
the covariant derivative can be rewritten as  
\begin{eqnarray}
D_{\mu} &=& 
\partial_{\mu} 
+ \frac{i e}{\sin \theta}  
\left[ T^{\pm}_{h} + T^{\pm}_{l} 
       + \lambda \sin^2 \phi \left( 
       \cos^2 \phi T^{\pm}_{h} - \sin^2 \phi T^{\pm}_{l} \right) 
\right] W^{\pm}_{\mu} 
\nonumber \\
&&+\frac{i e}{\sin \theta} 
\left[ \frac{\cos \phi}{\sin \phi} T^{\pm}_{h} 
       - \frac{\sin \phi}{\cos \phi} T^{\pm}_{l}
       - \lambda \sin^3 \phi \cos \phi 
         \left( T^{\pm}_{h} + T^{\pm}_{l} \right)
\right] W^{\prime \pm }_{\mu}   
\nonumber \\
&&+\frac{i e}{\sin \theta \cos \theta}  
\left[ T_{3 h} + T_{3 l} - Q \sin^2 \theta 
       + \lambda \sin^2 \phi 
       \left( \cos^2 \phi T_{3 h} - \sin^2 \phi T_{3 l} \right) 
\right] Z_{\mu}  
\nonumber \\ 
&&+\frac{i e}{\sin \theta} 
\left[ \frac{\cos \phi}{\sin \phi} T_{3 h} 
       - \frac{\sin \phi}{\cos \phi} T_{3 l} 
       - \lambda \frac{\sin^3 \phi \cos \phi}{\cos^2 \theta} 
       \left( T_{3 h} + T_{3 l} - Q \sin^2 \theta \right) 
\right] Z^{\prime}_{\mu} 
\nonumber \\
&&+ i e Q A_{\mu}~~.  
\end{eqnarray} 
Our model contains the FCNC interactions at tree level 
since the couplings of the third generation left--handed fermions 
to the $Z$, $ Z^{\prime}$ are different from those of 
the first and second generations.
We will discuss the FCNC and lepton mixing in detail in section IV.

\section{The Epsilon Variables in the Topflavour Model}

In the original work \cite{altarelli2}, three variables 
$\e_1$, $\e_2$ and $\e_3$ were defined from the basic observables,
the mass ratio of $W$ and $Z$ bosons $m_W/m_Z$, the leptonic
width $\Gamma_l$ and the forward--backward asymmetry for
charged leptons $A_{FB}^l$.
These observables are all defined at the $Z$--peak,
precisely measured and free from important
strong interaction effects such as $\alpha_s(m_Z)$ 
or the $Z \to b \bar b$ vertex.
In terms of these observables, $\e_1$, $\e_2$ and $\e_3$,
we have the virtue that the most interesting physical results are 
already obtained at a completely model independent manner
without assumptions like the dominance of vacuum polarization diagrams.

Because of the large $m_t$--dependent SM corrections to 
the $Z \to b \bar b$ vertex, however, 
the $\epsilon_i$'s and $\Gamma_b$ can only be 
correlated for a given value of $m_t$.
In order to overcome this limitation, Altarelli et al. \cite{altarelli1}
added a new parameter, $\epsilon_b$, which encodes the $m_t$--dependent
corrections to $Z \to b \bar b$ vertex and slightly modified 
other $\epsilon_i$'s.
Hence the four $\epsilon_i$'s are defined from an enlarged set
of basic observables $m_W/m_Z$, $\Gamma_l$, $A_{FB}^l$ and $\Gamma_b$
without need of specifying $m_t$.
Consequently the $m_t$--dependence for all observables
via loops come out through the $\epsilon_i$'s.
We work with this extended scheme here because we are interested in
the corrections to $Z \to b \bar b$ vertex.

The $\epsilon_{1,2,3}$ variables are defined by 
the linear combinations of
the correction terms $\Delta \rho$, $\Delta k$ and $\Delta r_{_W}$ 
in the eqs. (9) of ref. \cite{altarelli1}, 
which are extracted by the relations
\be
g_{_{lA}} &=& -\frac{1}{2} \left( 1+\frac{\Delta \rho}{2} \right)~~,
\nonumber 
\ee
\be
\frac{g_{_{lV}}}{g_{_{lA}}} &=& 1 - 4 (1+ \Delta k) s_0^2~~,
\nonumber 
\ee
\be
\left( 1- \frac{m_W^2}{m_Z^2} \right) \frac{m_W^2}{m_Z^2}
&=& \frac{\pi \alpha(m_Z)}
         {\sqrt{2}~ G_F~ m_Z^2 (1-\Delta r_{_W})}~~,
\ee
with
\be
s_0^2 c_0^2 = \frac{\pi \alpha(m_Z)}{\sqrt{2}~ G_F~ m_Z^2}~~,
\nonumber
\ee
and $c_0^2 = 1- s_0^2$.
The vector and axial--vector couplings for charged leptons,
$g_{_{lV}}$ and $g_{_{lA}}$ are obtained from the observables
$\Gamma_l$ and $A_{FB}^l$.
Note that the university of couplings is violated in this model
as a result of the different flavour dynamics of the third
generation.
For the consistency, we define $\Delta \rho$ and $\Delta k$
from the couplings of the first and second generations only.
Here, we use 
the electronic width, $\Gamma_e$ and the muon forward-backward
asymmetry, $ A^{\mu}_{FB}$ to make the bound on $\epsilon_i$
harder.
 
The formula for $\epsilon_b$ is rather complicated. 
It is defined by the equations :
\be
g_{_{bA}} = -\frac{1}{2} \left( 1+\frac{\Delta \rho}{2} \right)
          (1+\epsilon_b)~~,
\nonumber
\ee
\be
\frac{g_{_{bV}}}{g_{_{bA}}} 
     = \frac{ 1-\frac{4}{3}(1+\Delta k) s_0^2 + \epsilon_b}
            {1+\epsilon_b}~~.
\ee
We obtain the relation between $\epsilon_b$ and $\Gamma_b$ by 
insertion of $g_{_{bV}}$ and $g_{_{bA}}$ into the formula
of $\Gamma_b$.

The epsilon variables are obtained using the recent LEP data 
listed in table I taken from ref. \cite{LEPEWWG}:
\be
\epsilon_1 &=& (4.7 \pm 1.7) \times 10^{-3}~~,
\nonumber \\
\epsilon_2 &=& (-5.5 \pm 1.8) \times 10^{-3}~~,
\nonumber \\
\epsilon_3 &=& (4.3 \pm 2.5) \times 10^{-3}~~,
\nonumber \\
\epsilon_b &=& (-2.6 \pm 2.0) \times 10^{-3}~~.
\ee

The basic observables in the topflavour model depend upon
the model parameters $\lambda$ and $\sin^2 \phi$
as well as the Higgs mass $m_{_H}$.
The vector and axial--vector couplings of fermions
to $Z$ boson at tree level are given by
\be
v_f &=& T_{3h} + T_{3l} - 2 Q_f \sin^2 \theta
      + \lambda \sin^2 \phi ( T_{3h} \cos^2 \phi - T_{3l} \sin^2 \phi )~~,
\nonumber \\
a_f &=& T_{3h} + T_{3l} 
      + \lambda \sin^2 \phi ( T_{3h} \cos^2 \phi - T_{3l} \sin^2 \phi )~~,
\ee
where $\sin^2 \theta$ is defined as the shifted Weinberg angle at tree level
by the new physics effects
\be
\sin^2 \theta = s^2_0 - \lambda \sin^4 \phi \frac{c_0^2 s_0^2}
                                           { c_0^2 - s_0^2 }~~,
\ee
up to the linear order of $\lambda$.
We include the radiative corrections and obtain 
the effective fermion couplings and Weinberg angle 
by introducing the electroweak form factor 
as done in ref. \cite{ZFITTER}: 
$\sin^2 \theta^f_{eff} = \kappa_f \sin^2 \theta$
and $g_A = \sqrt{\rho_f} a_f$.
Along with the eqs. (7), (8) and (11), the effective fermion couplings 
including new physics corrections for lepton and $b$--quark 
define the correction terms 
$\Delta \rho_{new}$, $\Delta k_{new}$, 
and $\epsilon_b^{new}$.
The new physics contribution to the mass of $W$ boson are encoded
in the effective Weinberg angle and also define the term 
$\Delta r_{_W}^{new}$ :
\be 
\sin^2 \theta_{eff} \cos^2 \theta_{eff}
=\frac{\pi \alpha(m_Z)}{\sqrt{2}~ G_F~ m_Z^2 (1-\Delta r_{_W}^{new})}~~.
\ee
With the correction terms we obtain the epsilon variables
in the topflavour model :
\be
\epsilon_1 &=& \epsilon^{SM}_1 - 2 \lambda \sin^4 \phi
\nonumber \\
\epsilon_2 &=& \epsilon^{SM}_2 +  \lambda \sin^4 \phi 
                             \left( \frac{s_0^2}{c_0^2 - s_0^2} -2
                             \right)
\nonumber \\
\epsilon_3 &=& \epsilon^{SM}_3 - \lambda \sin^4 \phi
\nonumber \\
\epsilon_b &=& \epsilon^{SM}_b - \frac{1}{2} \lambda 
                        \frac{\sin^2 \phi}{g_{lA}^{SM}}
                        (1+\epsilon_b^{SM} \sin^2 \phi)
\ee
(dropping the tag "new" from now on).
In the LEP era, it is natural to use the precisely
measured values of $\alpha(m_{_Z})$, $G_F$, $M_{_Z}$.
We used the ZFITTER \cite{ZFITTER} for numerical
calculations of the epsilon variables in this model.
We use 175 GeV as input value of $m_t$, 
which is the central value
of the recent CDF and D0 report \cite{top}.

%
In Figs. 1 and 2, the experimental ellipses for 1-$\sigma$ level and
90\%, 95\% confidence levels are shown in the
$\epsilon_1-\epsilon_b$ 
and $\epsilon_3-\epsilon_b$ planes respectively
with the results of the SM and the topflavour model. 
We express the results with variations of
$m_H$, $\sin^2 \phi$ and $m_{_{Z'}}$.
Note that the mass of the neutral heavy gauge boson $Z'$
is used instead of the parameter $\lambda$ to be comprehensive
because  $m_{_{Z'}}$  is an observable.
The mass of $Z'$ is related to $\lambda$ as follows :
\be
m_{Z'}^2 = \frac{m_Z^2}{\lambda \sin^2 \phi \cos^2 \phi}
           \left[ \cos^2 \theta_{SM} + \lambda \sin^4 \phi
                  \left( \frac{c_0^2 s_0^2}{c_0^2-s_0^2} + 2 c_0^2
                  \right)
           \right]~~,
\ee
where $\theta_{SM}$ is the Weinberg angle of the SM.
The ellipses tend to be shifted along the minus direction of
$\epsilon_b$ compared with those from 1996 data \cite{kylee,parklee}
and thus come closer to the SM prediction.
We find that the SM predictions lie
inside the 90 \% C.L. ellipses in both of
$\epsilon_1-\epsilon_b$ and $\epsilon_3-\epsilon_b$ planes.
The remaining deviations are caused by the $Z b \bar{b}$ coupling
and related to the $R_b$ discrepancy.
In Fig. 3 we show the parameter space of $\sin^2 \phi - m_{_{Z'}}$
and allowed region constrained by $\epsilon_{1,3,b}$.
For small $\sin^2 \phi$, $\epsilon_b$ give the most strict
constraint but constraint of $\epsilon_1$ becomes more
important when $\sin^2 \phi > 0.32$.
This arises from the shift of the electroweak couplings of leptons
because the $SU(2)_l$ coupling constant $g_l$ is rather large 
corresponding to the large mixing angle $\phi$.
The lower limit of the mass of heavy gauge boson $Z'$
is about 1.2 TeV (1.1 TeV) at 90 (95) \% C. L.
which agrees with the result of ref. \cite{malkawi}.

\section{FCNC and Lepton Mixings}
     
The FCNC processes violating lepton family number 
are possible in this model 
since the couplings of neutral currents for the third fermions to
$Z$ and $Z'$ gauge bosons are different from those of the
first and second generations.
The neutral current interactions with $ Z$ and $ Z'$ 
can be rewritten as universal part and nonuniversal part to the
third generations separately.
\begin{eqnarray} 
 {\cal L}_0 &=&  {\cal L}_{I} + {\cal L}_{3}~~, 
\ee
where the universal part is given by
\be
 {\cal L}_{I} &=&  {\sum_{F=U,D,E,N}}  
    {\bar F}_L {\gamma}_{\mu}  
       \left[ 
               G^{(F)}_{L} Z^{\mu}  + G^{\prime (F)}_{L} Z^{\prime \mu} 
       \right] 
         F_L     
\nonumber \\
  &&~~~~~+
     {\bar F}_R {\gamma}_{\mu} 
       \left[ 
               G^{(F)}_{R} Z^{\mu}  + G^{\prime (F)}_{R} Z^{\prime \mu} 
       \right]  F_R~~,
\ee
and the nonuniversal part by
\be
  {\cal L}_{3} &=&  
     {\bar F}_L {\gamma}_{\mu} 
     \left[ 
            X^{(F)}_{L} Z^{\mu}  + X^{\prime (F)}_{L} Z^{\prime \mu} 
     \right] 
      F_L ~~,
\ee
where $U \equiv ( u_0 , c_0 , t_0 ) , D \equiv  ( d_0 , s_0 , b_0 ) ,
 E \equiv ( e_0  ,  \mu_0   , \tau_0  ) $ and 
$ N  \equiv ( \nu_{e0} , \nu_{\mu 0} , \nu_{\tau 0} ) $
are electroweak states.
The couplings of fermions to $ Z $ and $ Z^{\prime}$ are given by
\begin{eqnarray}
G^{F}_L  &=&  - \frac{e}{\cos \theta \sin \theta} 
              \left[
              T_{3 F} - Q_F \sin^2 \theta 
              -  \lambda \sin^4 \phi T_{3 F}  
                                      \right]  I 
\nonumber \\
G^{F}_R  &=& \frac{e}{\cos \theta \sin \theta} 
              \left[ 
               Q_F \sin^2 \theta 
              \right]  I
\nonumber \\
G^{\prime F}_L  &=& \frac{e}{\sin \theta} 
                  \left[
                  \frac{ \sin \phi}{\cos \phi} T_{3 F}
                  +\lambda \frac{ \sin^3 \phi \cos \phi }
                                { \cos^2 \theta} 
                  \left( T_{3 F} - Q_F \sin^2 \theta \right) 
                  \right]  I 
\nonumber \\
G^{\prime F}_R &=& - \frac{e}{\sin \theta}
              \left[
              \lambda Q_F \tan^2 \theta \sin^3 \phi \cos \phi    
                    \right]  I
\end{eqnarray} 
where $I$ is the $3 \times 3$ identity matrix
and
\begin{eqnarray} 
 X^F_L        & = & -\frac{e}{\cos \theta \sin \theta} 
               \left( \begin{array}{ccc}
                0 &    0 &    0 \\
                0 &    0 &    0 \\
                0 &    0 &    \lambda \sin^2 \phi T_{3 F} \\
                            \end{array} 
                       \right) , \mbox{~~   }  
\nonumber \\
X^{\prime F }_L & = & - \frac{e}{\sin \theta} 
               \left( \begin{array}{ccc}
                0 &    0 &    0 \\
                0 &    0 &    0 \\
                0 &    0 &    \frac{1}{\sin \phi \cos \phi } T_{3 F} \\
                            \end{array}
                          \right)~~.
\end{eqnarray}
Therefore, the neutral current interactions are not
simultaneously diagonalized with the charged lepton mass
matrix by a unitary transformation, and the FCNC 
interaction terms are generated from the eq. (19).

Since the $e - \mu$ mixing process should be highly suppressed
by the experimental bound 
Br$(\mu^- \to e^- e^+ e^-) < 1.0 \times 10^{-12}$ \cite{pdg},
we consider two cases of $\mu - \tau$ and $e - \tau$ mixings here.
Introducing the new parameter $\sin \beta $ for $\mu - \tau$  
mixing, we obtain the mass eigenstates through the mixing matrix
\begin{equation}
\left( \begin{array}{c}  
       e \\
       \mu  \\
       \tau   \\
       \end{array} 
\right)_L  =
   \left( \begin{array}{ccc}
         1     &        0        &     0        \\
         0     &    \cos \beta   &   -\sin \beta  \\
         0     &    \sin \beta   &    \cos \beta  \\
        \end{array} 
   \right)
   \left( \begin{array}{c}  
       e_0 \\
       \mu_0  \\
       \tau_0   \\
       \end{array} 
    \right)_{L} ~~.
\end{equation}      
Thus ${\cal L}_3$ yields the FCNC interaction terms 
\begin{equation}
 {\cal L}_{int} = 
    \lambda \sin \beta \cos \beta 
     \left( 
          \frac{e}{\cos \theta \sin \theta}
          \frac{ \sin^2 \phi}{2} 
     \right) 
      {\bar \tau}_L {\gamma}_{\mu}  \mu_L Z^{\mu}  
      +  h.c. 
\end{equation}
through which $ \tau^- \rightarrow \mu^- \mu^- \mu^+  $
and $Z \rightarrow \mu^{\pm} \tau^{\mp} $ decays are possible.
We express their branching ratio with model parameters:
\be
\frac{\mbox{Br}(\tau^- \to \mu^- \mu^- \mu^+)} 
     {\mbox{Br}(\tau^- \to \bar{\nu}_{\mu} \mu^- \nu_{\tau})}
&=& 0.262
            \lambda^2 \sin^2 \beta \cos^2 \beta
            \left[ \frac{1}{2} 
              (\sin^2 \beta - 4 \sin^2 \theta \sin^2 \phi)^2
              + \sin^4 \theta \sin^4 \phi
            \right]
\nonumber \\
\mbox{Br}(Z \to \mu^{\mp} \tau^{\pm}) &=&
      \frac{0.332}{(\Gamma_Z/1~\mbox{GeV})}
            \lambda^2 \sin^4 \phi \sin^2 \beta \cos^2 \beta 
 ~~.
\ee
We can constrain the parameter space by 
the experimental bounds \cite{pdg} :
\be
\mbox{Br}(\tau^- \to \mu^- \mu^- \mu^+) < 1.9 \times 10^{-6}
&&~~~\mbox{(at 90 \% C.L.)} 
\nonumber \\
\mbox{Br}(Z \to \mu^- \tau^+) < 1.7 \times 10^{-5}
&&~~~\mbox{(at 95 \% C.L.)} ~~.
\ee
In Fig. 4 (a) we show the allowed region in the 
$\sin^2 \phi$--$m_{_{Z'}}$ plane by the experimental bound
of eq. (23)
together with $\epsilon_b$ bound.
We find that in the region of small $\sin^2 \phi$
the additional constraints on the allowed region
can be given from the experimental bound
of $\tau \to \mu \mu \mu$ 
when the value of $\sin \beta$ is large enough.
In the most region of the parameter space, however,
these experimental bounds are not so restrictive
compared to the LEP experimental bounds
and there are no limits on $\sin \beta$.

If we assume the $(e - \tau)$ mixing,
another mixing angle $\gamma$ is introduced through
the mixing matrix
and we have the FCNC term
\begin{equation}
 {\cal L}_{int} = 
    \lambda \sin \gamma \cos \gamma 
     \left( 
         \frac{e }{\cos \theta \sin \theta} 
         \frac{ \sin^2 \phi}{2} 
     \right) 
      {\bar \tau}_L {\gamma}_{\mu}  e_L Z^{\mu}  
      +  h.c. ~~.
\end{equation}
We also show the allowed parameter space in
$\sin^2 \gamma$--$m_{_{Z'}}$ plane in Fig. 4 (b)
using the experimental bound
Br$(\tau^- \to e^- e^- e^+) < 3.3 \times 10^{-6}$
(at 90 \% C.L.) and
Br$(Z \to e^- \tau^+) < 9.8 \times 10^{-6}$
(at 95 \% C.L.) \cite{pdg}. 
Constraints on the allowed parameter space are similar
to the case of $(\mu - \tau)$ mixing.

\section{Limits from the Atomic Parity Violation Experiments}

The parity violating processes in heavy atoms are known to
provide an excellent test for the electroweak theory
at the energy scale far below $m_{_Z}$.
Especially for the extended gauge interactions, 
stringent constraints on the mass of extra gauge bosons
can arise from the atomic parity violation experiments.
At low energies, the parity formula electron--quark
interactions are described by the effective lagrangian :
\be
{\cal L}_{eff} = \frac{G_F}{\sqrt{2}} 
              (\bar{e} \gamma_{\mu} \gamma_5 e)
              \left( C_{1u} \bar{u} \gamma^{\mu} u
                   + C_{1d} \bar{d} \gamma^{\mu} d \right)~~.
\ee
The parity violating amplitudes are expressed by 
the so-called weak charge $Q_W$ of the atom
\be
Q_W = -2 \big[ C_{1u} (2Z+N) + C_{1d} (Z+2N) \big]~~,
\ee
where $Z$ is the number of protons and $N$ the number of neutrons.

In the topflavour model, the effective lagrangian of eq. (25) is obtained
by integrating out the $Z$ and $Z'$ bosons with the coefficients
$C_{1u}$ and $C_{1d}$
\be
C_{1u,1d} = C_{1u,1d}^{SM} ( 1- \lambda \sin^4 \phi )
\ee
in terms of the model parameters $\lambda$ and $\sin^2 \phi$
up to the leading correction of $\lambda$.
The shift in $Q_W$ away from the SM prediction by the
new physics effects are given by
\be
\Delta Q_W = Q_W - Q_W^{SM} = - Q_W^{SM} \lambda \sin^4 \phi~~.
\ee

With the recent accurate measurements of the APV in $Cs$ atom
\cite{wood,dzuba}, one obtain
\be
Q_W(^{133}_{55} Cs) = -72.41 \pm 0.25 \pm 0.80~~,
\ee
where the first error comes from experiment and the second from
atomic theory.
From the SM prediction \cite{marciano}
\be
Q_W^{SM}(^{133}_{55} Cs) = -73.20 \pm 0.13 
\ee
for a Higgs boson mass of 100 GeV, 
the deviation $\Delta Q_W$ is estimated as
\be
\Delta Q_W(^{133}_{55} Cs) = 0.79 \pm 1.06~~,
\ee
in which the error includes both of theoretical and experimental ones.
We obtain the constraint on the parameter space
\be
\lambda \sin^4 \phi \le - \frac{\Delta Q_W}{Q_W^{SM}} 
= 0.0108 \pm 0.0145~~,
\ee
which is already shown in Fig. 3.
The lower limit of the $Z'$ boson mass from the APV 
is found to be
\be
m_{_{Z'}} > 480~~ \mbox{GeV}~~.
\ee
We find that the APV experiment does not give additional
constraints on the allowed parameter space that can compete with
the LEP data at $Z$--peak.

\section{Concluding Remarks}

In this work, we explore the phenomenologies of the topflavour model, 
which are extension of the SM with the additional $SU(2)$ symmetry.
In this model, the third generation is special
and it is expected to explain the hierarchy of the fermion
mass spectrum as well as the $R_b$ discrepancy observed
in LEP experiment.
In this paper we first analyzed the model in terms of the electroweak
precision variable $\epsilon_{1,2,3,b}$ and critically constrained
the parameter space of the model.
Because the topflavour model has nonstandard interactions 
on the lepton sector as well as the quark sector, 
it would be reasonable to use 
the electroweak precision variable to test the model.
The experimental values of $\epsilon$ variables are obtained
from the recent LEP results reported 
by the LEP Electroweak Working Group.
At present the SM predictions is out of 1--$\sigma$ ellipses 
mainly caused by the deviations of $\epsilon_b$
originated by the $Z b \bar{b}$ vertex.
Since the topflavour model provides the different flavour dynamics
on the third generations, the value of $\epsilon_b$ can shift
to the experimental value.
We did not present the result in the $\epsilon_2- \epsilon_b$
plane because we used the value of the $W$ boson mass 
fitted to LEP data alone in ref. \cite{LEPEWWG}.
If we use the directly measured data, errors on $m_{_W}$
is still too large to give meaningful constraint.
As the more precise value of the $W$ boson mass is obtained,
$\epsilon_2$ variable can also provide a stringent test
for the theoretical predictions.
The characteristic feature of this model is the violation 
of the universality in the interactions of neutral currents
which result in the dangerous FCNC even in the lepton sectors. 
Thus the parameter space is additionally constrained 
by the experimental bounds on lepton family number changing processes.
For most values of $\sin \beta$ ($\sin \gamma$),
$\epsilon_i$'s give more stringent constraints
except for the region where $\sin^2 \phi$ is small.
The APV experiment does not give a stringent restriction
on this model compared with $Z$-peak data. 

In conclusion, we investigated the phenomenological implications
of the topflavour model with the LEP data at $Z$--peak and
the APV data at low energy scale.
We found that the best fitted mass of $Z'$ boson to the LEP data 
is about 2 TeV and with this value, the signature of $Z'$ boson
is expected to be found via the excess in the $t \bar{t}$
pair production at the LHC or at the NLC, as is discussed
in ref. \cite{malkawi}.

\begin{center}
{\bf Acknowledgement}
\end{center}
\vskip 1.0cm

We thank Prof. P. Ko for careful reading of the manuscript
and valuable comments.
This work was supported in part by 
the Korean Science and Engineering Foundation (KOSEF).

\begin{table}
\begin{center}
\caption{
The LEP data reported by the LEP Electroweak Working Group
in ref [1].
}
\label{Table1}
\vspace{2cm}
\begin{tabular}{|clc|ccc|}
 & $M_W$ & & & 80.298 $\pm$ 0.043 GeV& \\
 & $M_Z$ & & & 91.1867 $\pm$ 0.0020 GeV& \\
 & $\Gamma_e$ & & & 83.94 $\pm$ 0.14 MeV& \\
 & $A_{FB}^{\mu}$ & & & 0.0163 $\pm$ 0.0014 & \\
 & $R_b$ & & & 0.2174 $\pm$ 0.0009 & \\
 & $\Gamma_{had}$ & & & 1743.2 $\pm$ 2.3 MeV& \\
\end{tabular}
\end{center}
\end{table}

%

\newpage

\vskip 1.0cm
{\large \bf Figure Captions}
\vskip 2.0cm

{\bf Fig. 1}\\
Plots of the model predictions in units of $10^{-3}$
with varying the model parameter
$sin^2 \phi$, $m_{_{Z'}}$ and the Higgs boson mass $m_H$ in
$\epsilon_1$--$\epsilon_b$ plane.
The experimental ellipses at 1-$\sigma$, 90 \% C.L.
and 95 \% C.L. are given.
Points are labelled from $\sin^2 \phi = 0.1~ \sim ~0.9$
from right to left for each value of $m_{_{Z'}}$.
\vskip 1.0cm

{\bf Fig. 2}\\
Plots of the model predictions in units of $10^{-3}$
with varying the model parameter
$sin^2 \phi$, $m_{_{Z'}}$ and the Higgs boson mass $m_H$ in
$\epsilon_3$--$\epsilon_b$ plane.
\vskip 1.0cm

{\bf Fig. 3}\\
The parameter space of $\sin^2 \phi - m_{_{Z'}}$ plane
constrained by $\epsilon_1$, $\epsilon_3$, $\epsilon_b$
and $\Delta Q_W$ from the atomic parity violation.
The Higgs boson mass $m_H = 100$ GeV and the top quark
mass $m_t = 175$ GeV.
For $\epsilon_i$'s, the solid lines denote the lower limit
of 90\% C.L. and the dashed line of 95 \% C. L..
The region above each curve is allowed.
\vskip 1.0cm

{\bf Fig. 4}\\
(a) The parameter space of $\sin^2 \phi - m_{_{Z'}}$ plane
constrained by $\epsilon_b$ and lepton family number
violating processes $\tau \to \mu \mu \mu$, $Z \to \tau \mu$
with $(\mu - \tau)$ mixing.
Increasing curves when $\sin^2 \phi$ goes to 0 come from
the experimental bound of $\tau \to \mu \mu \mu$
and decreasing curves from the bound of $Z \to \tau \mu$.
(b) The parameter space of $\sin^2 \phi - m_{_{Z'}}$ plane
constrained by $\epsilon_b$ and lepton family number
violating processes $\tau \to e e e $, $Z \to \tau \tau e$
with $(e - \tau)$ mixing.
Increasing curves when $\sin^2 \phi$ goes to 0 come from
the experimental bound of $\tau \to e e e$
and decreasing curves from the bound of $Z \to \tau e$.

\newpage

\begin{figure}[bh]
\label{figone}
\caption{}
\centering
\centerline{\epsfig{file=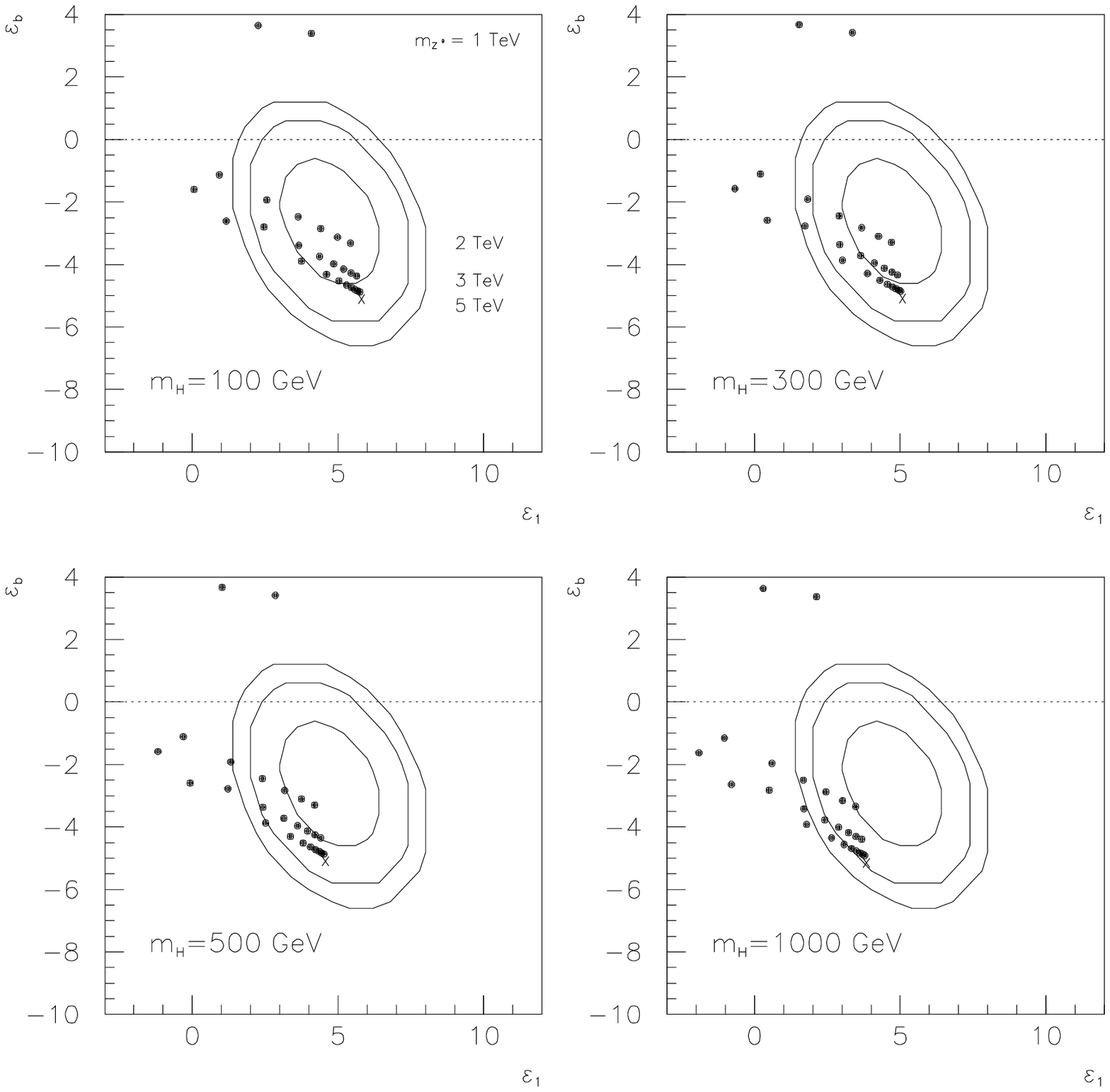}}
\end{figure}

\newpage
\begin{figure}[bh]
\label{figtwo}
\caption{}
\centering
\centerline{\epsfig{file=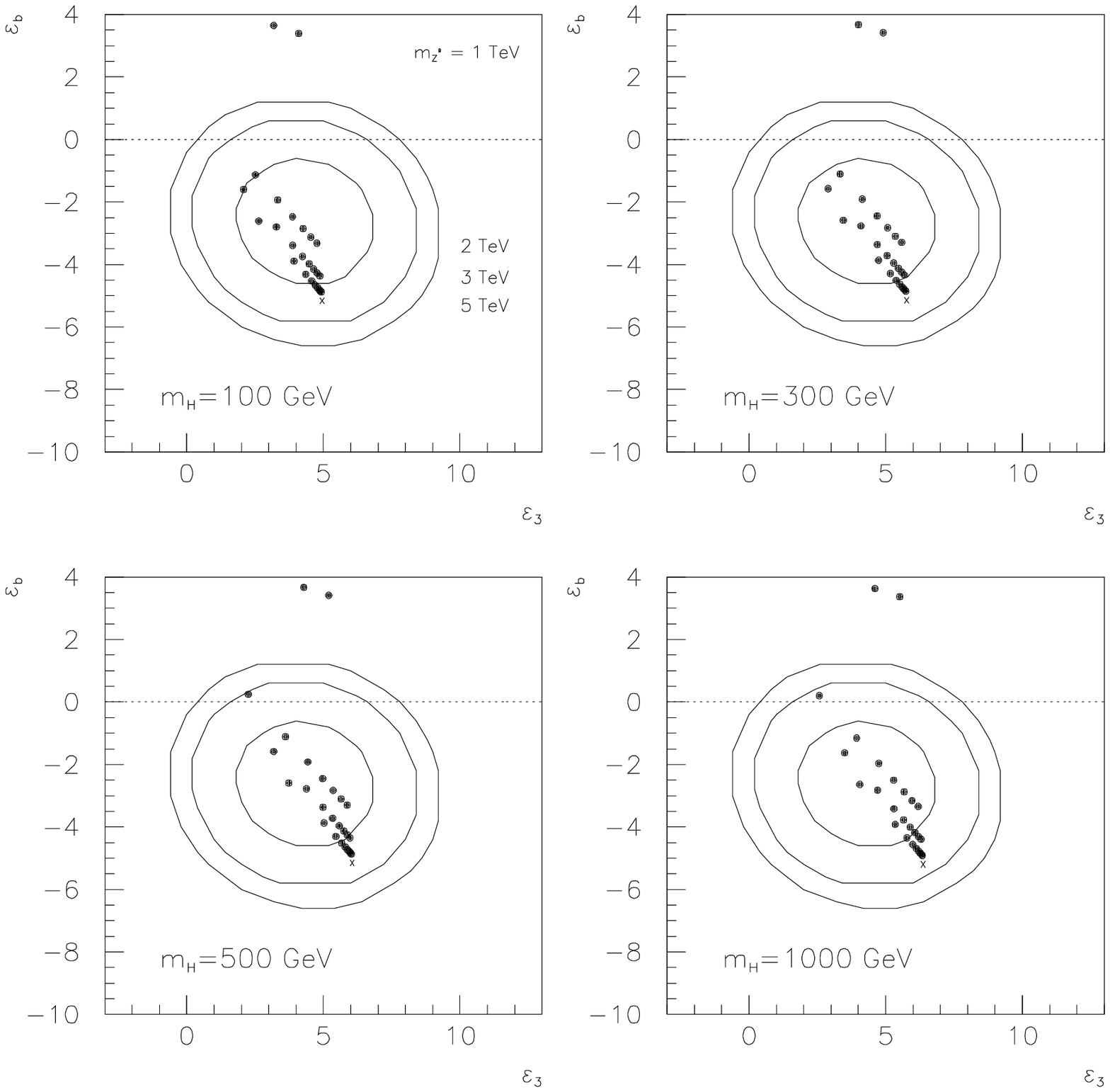}}
\end{figure}

\newpage
\begin{figure}[bh]
\label{figthree}
\caption{}
\centering
\centerline{\epsfig{file=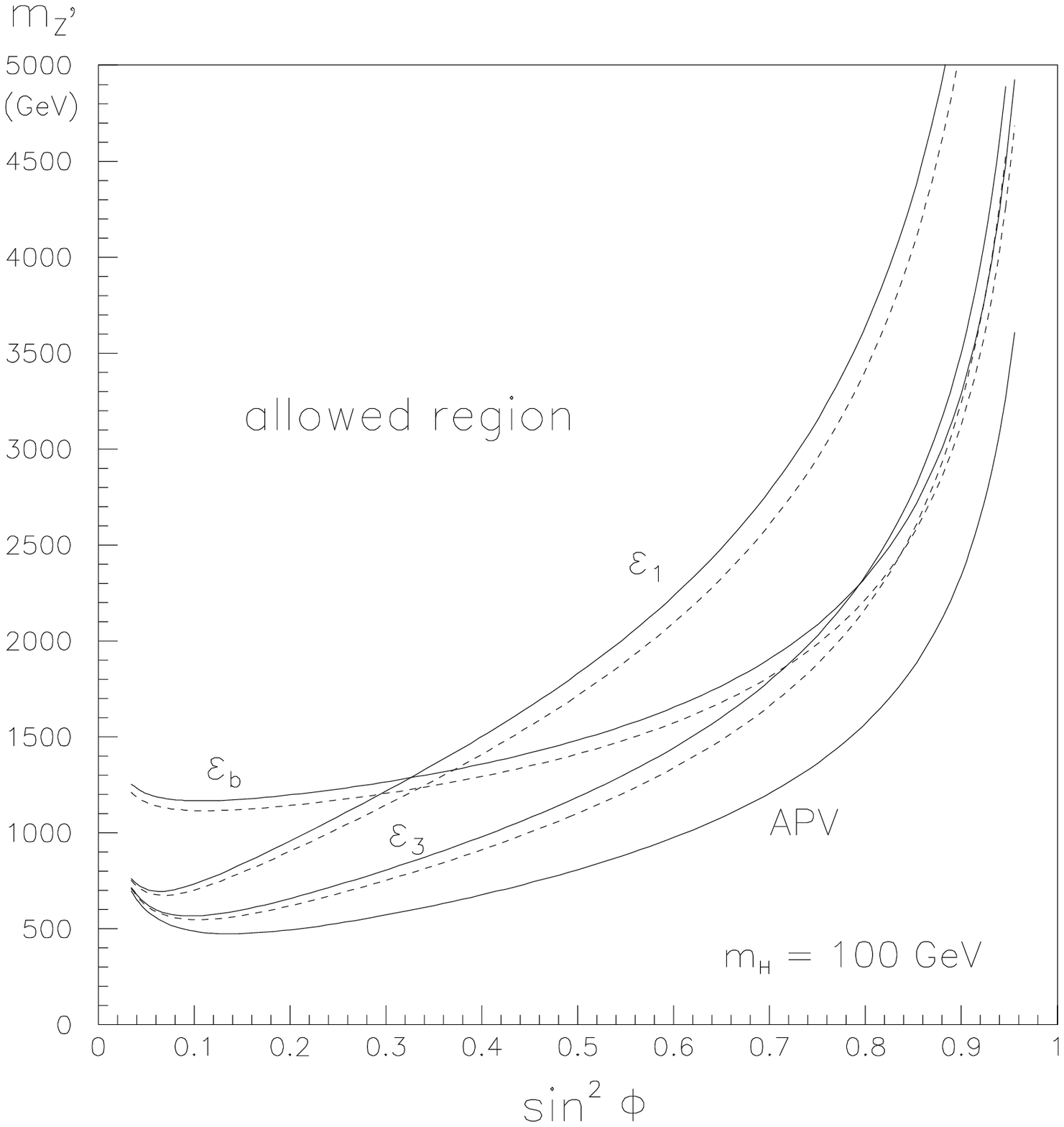}}
\end{figure}

\newpage
\begin{figure}[bh]
\label{figfoura}
\caption{}
\centering
\centerline{\epsfig{file=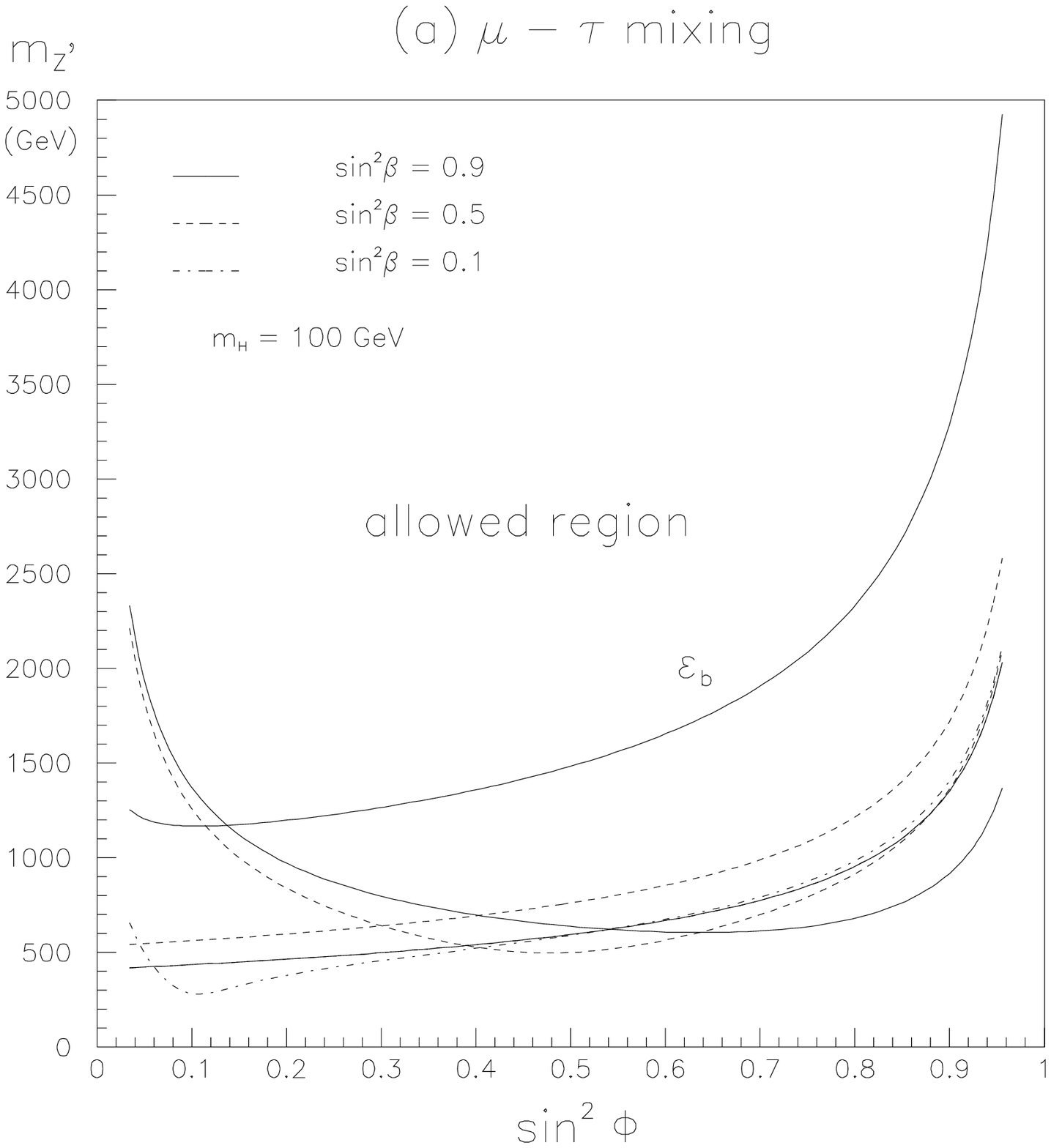}}
\end{figure}

\newpage
\begin{figure}[bh]
\label{figfourb}
\caption{}
\centering
\centerline{\epsfig{file=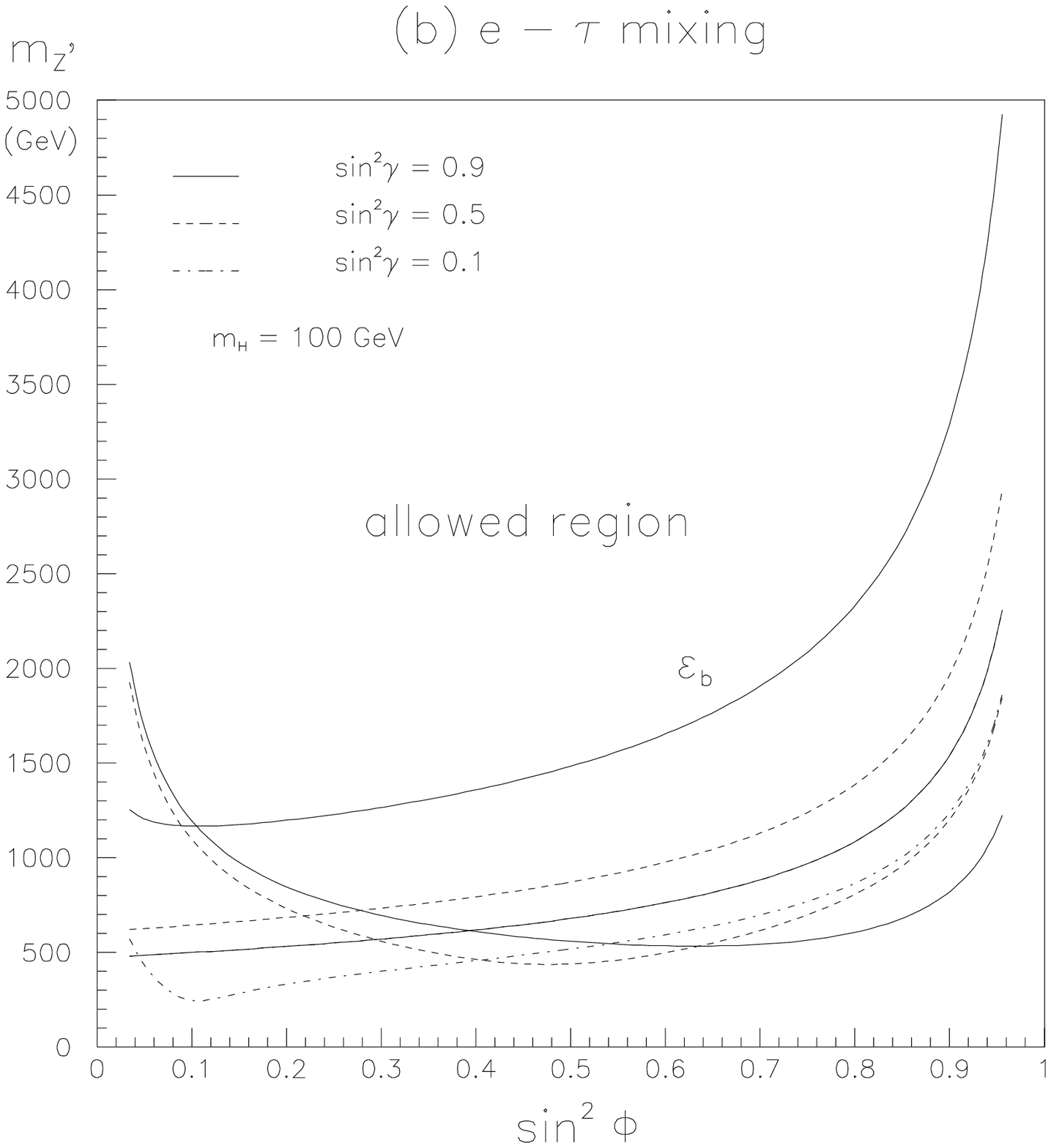}}
\end{figure}

\end{document}